\documentclass[12pt,referee,pdflatex,sn-vancouver-num]{sn-jnl}

\usepackage{graphicx}%
\usepackage{multirow}%
\usepackage{amsmath,amssymb,amsfonts}%
\usepackage{amsthm}%
\usepackage{mathrsfs}%
\usepackage[title]{appendix}%
\usepackage{xcolor}%
\usepackage{textcomp}%
\usepackage{manyfoot}%
\usepackage{booktabs}%
\usepackage{algorithm}%
\usepackage{algorithmicx}%
\usepackage{algpseudocode}%
\usepackage{listings}%
\usepackage{adjustbox}
\usepackage{subfigure}
\usepackage{array} 
\usepackage{tabularx}
\usepackage{makecell}

\theoremstyle{thmstyleone}%
\theoremstyle{thmstyletwo}%

\theoremstyle{thmstylethree}%

\begin{document}

\newgeometry{
left=1in,
right=1in,
top=1in,
bottom=1in
}

\title[Article Title]{Abdominal Ultrasound Simulation from Semantic Labels using Paired Label-to-Physics-Based Image Translation}


\author*[1,2]{\fnm{Santiago} \sur{Vitale}}\email{santiago.vitale@pladema.exa.unicen.edu.ar}

\author[1,2]{\fnm{Duilio} \sur{Deangeli}}\email{ddeangeli@pladema.exa.unicen.edu.ar}
\equalcont{These authors contributed equally to this work.}

\author[1,2]{\fnm{Ignacio}\sur{Larrabide}}
\author[1,2]{\fnm{José~Ignacio}\sur{Orlando}}

\affil*[1]{\orgname{National Scientific and Technical Research Council (CONICET)}, \orgaddress{\street{Godoy Cruz 2290}, \city{CABA}, \postcode{C1425FQB}, \state{Buenos Aires}, \country{Argentina}}}

\affil[2]{\orgdiv{Pladema}, \orgname{UNICEN}, \orgaddress{\street{Pasaje Arroyo Seco}, \city{Tandil}, \postcode{7000}, \state{Buenos Aires}, \country{Argentina}}}


\abstract{
\textbf{Purpose:} Current abdominal ultrasound (US) simulation methods often require CT-based anatomical references for ray-casting, limiting deformation and pathology variability. We propose a learning-based pipeline trained to predict physics-based images derived from CT scans from semantic labels, enabling controlled simulations without patient-specific CT volumes at inference time.

\textbf{Methods:} We introduce a two-stage pipeline that maps anatomical segmentations to realistic US images through a \textcolor{black}{simplified US image}. Stage~I synthesizes this image from semantic labels using models trained on CT-based ray-casting outputs. Stage~II refines it into a realistic US scan using anatomically guided unpaired translation. Deformations and pathologies are generated by editing anatomical maps.

\textbf{Results:} We evaluated Pix2Pix and the Semantic Diffusion Model (SDM) in Stage~I, followed by segmentation-guided CycleGAN (SG-CycleGAN) refinement in Stage~II. SDM significantly outperformed Pix2Pix in morphological metrics, including MAE$\downarrow$ (19.85 vs. 21.65), SSIM$\uparrow$ (0.28 vs. 0.24), and mIoU$\uparrow$ (0.43 vs. 0.29), whereas Pix2Pix yielded better perceptual point estimates (LPIPS$\downarrow$: 0.17 vs. 0.19; FID$\downarrow$: 0.32 vs. 0.37; KID$\downarrow$: 0.25 vs. 0.48).

\textbf{Conclusion:} Training paired generative models with physics-based supervision enables approximation of CT-derived ray-casting outputs at inference time directly from semantic labels. \textcolor{black}{Although the pipeline does not require patient-specific CT volumes at inference time, CT-derived segmentations and ray-casting simulations remain necessary to train Stage~I. Once trained, the framework enables controllable healthy and pathological simulations through semantic-map modification.}}

\keywords{Generative models, Semantic Image Synthesis, US-simulation, Pathology simulation}


\maketitle

\section{Introduction}
\label{sec:intro}

Abdominal US is a non-invasive, real-time imaging technique widely used to assess abdominal structures and support clinical decision-making~\cite{rumack2017diagnostic,kameda2016overview}. However, its diagnostic accuracy heavily depends on the sonographer’s expertise~\cite{wang2022task}, which requires significant hands-on practice and access to a wide range of case scenarios to be developed, particularly for low-prevalence conditions. US simulators have emerged as effective training tools, offering controlled, repeatable environments where learners can acquire and refine scanning techniques without relying on patient availability~\cite{nayahangan2021simulation, hoffmann2020medical}. This approach has been extensively explored as an alternative e.g. for training radiologists in capturing and interpreting scans~\cite{dinh2016integration, ostergaard2016simulation}. Nonetheless, designing realistic and adaptable simulators remains a key challenge.

A common approach to US simulation involves modeling acoustic wave propagation through 3D volumes derived from CT or MRI scans~\cite{shams2008real}. While this enables anatomically faithful simulation, it introduces a critical dependence on patient-specific volumetric data. This restricts the ability to model dynamic phenomena (e.g., breathing or pressure deformations) or simulate rare pathological cases, especially when the corresponding CT data are unavailable. Several works have addressed this by applying ray-tracing rendering techniques over CT-derived segmentations~\cite{burger2012real, cambet2020high}. These methods reduce part of the dependence on raw intensity volumes by approximating wave propagation using hand-crafted mechanoacoustic tissue properties~\cite{cambet2020high}, but they may limit the realism and accuracy of simulated US artifacts~\cite{burger2012real}.
An alternative is the direct generation of images from semantic labels, a process frequently known as Semantic Image Synthesis (SIS)~\cite{wang2022semantic,ko2024stochastic}. GAN-based SIS models have been applied to generate localized anatomical structures such as ovaries~\cite{liang2022sketch} and kidneys~\cite{Pigeau}. 

More recently, diffusion models have emerged to improve the realism and controllability of US synthesis. Denoising Diffusion Probabilistic Models (DDPMs)~\cite{ho2020denoising}, for instance, have been used for cardiac US synthesis guided by label maps~\cite{stojanovski2023echo} and thyroid US generation from healthy volunteers~\cite{dominguez2024diffusion}. More general conditional diffusion frameworks, such as ControlNet~\cite{zhang2023adding}, have shown that generative models can be guided by spatial conditions, including edges, sketches, depth maps, and semantic layouts. This idea is relevant for medical image synthesis because it provides a flexible mechanism for controlling image structure during generation. In US, related conditional or controllable synthesis approaches have been explored for specific organs or restricted anatomical settings, such as fetal anatomy-guided synthesis~\cite{duan2025fetalflex} and breast US generation with segmentation-based ControlNet conditioning~\cite{freiche2025ultrasound}. However, these works do not directly address multi-organ abdominal US simulation from dense anatomical masks.  

Despite this progress, semantic image synthesis for multi-organ abdominal US remains under-explored. Existing abdominal US synthesis approaches have mainly focused on CT-to-US translation, simulated-to-real refinement, or enhancement of physics-based simulations rather than direct generation from semantic anatomical masks. Zhang \textit{et al.}~\cite{zhang2020deep}, for example, used adversarial image translation to enhance low-quality simulated US images while preserving anatomical and acoustic information. Content-preserving unpaired translation approaches have also been proposed to reduce the appearance gap between simulated and real US images while maintaining the simulated anatomical layout~\cite{Tomar2021}. Similarly, Vitale \textit{et al.}~\cite{vitale2020improving} used CycleGAN-based refinement to improve the realism of abdominal US simulations generated from CT-derived ray-casting. More recent CT-to-US approaches, such as S-CycleGAN~\cite{song2024s} and domain-constrained CT-to-US synthesis~\cite{wang2025ct}, also aim to preserve anatomical structure during cross-domain translation.

These existing abdominal approaches rely either on CT images, CT-derived simulations, or precomputed simulated images as the input representation. They therefore remain constrained by the availability of patient-specific volumetric data or by the realism of the underlying physical simulator. Extending controllable synthesis to abdominal US is also challenging because this domain involves large fields of view, multiple interacting organs, acoustic shadowing, occlusions, view-dependent artifacts, and high anatomical variability. In addition, the scarcity of paired abdominal datasets with dense semantic labels and corresponding realistic US images limits direct supervised semantic synthesis. This motivates this work, where we propose a learning-based semantic image synthesis framework that generates abdominal US from anatomical segmentation masks. During training, ray-casting simulations obtained from CT scans are used to provide paired supervision signals; once trained, the proposed model performs image synthesis directly from semantic anatomical masks. Our pipeline combines \textcolor{black}{a new Stage~I SIS component with the SG-CycleGAN refinement strategy previously introduced in~\cite{vitaleimproving}. Stage~II is not redesigned in this work; instead, it is used as a downstream refinement module to evaluate whether the new Stage~I outputs remain compatible with the previously established simulated-to-real translation framework. The practical role of the proposed approach is to provide a semantic-label-driven alternative to the CT-based ray-casting step, allowing new anatomical configurations and proof-of-concept pathological scenarios to be generated from modified label maps without requiring a new patient-specific CT or MRI volume at inference time. This capability could support case generation for abdominal US simulators.}

\begin{figure*}[t]
    \centering
    \includegraphics[width=1\textwidth]{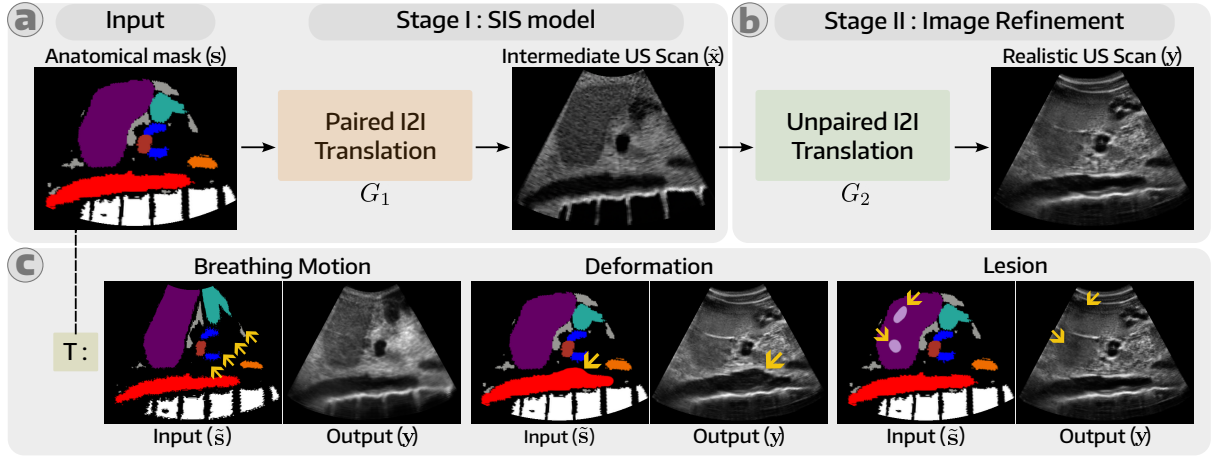}
    \caption{Overview of the proposed inference pipeline. (a) A semantic label map is translated into a simplified US image via a paired model. (b) An unpaired SG-CycleGAN refines realism while preserving anatomy. (c) Optional label transformations $T$ (e.g., motion, lesions, deformations) enable diverse scenario generation without CT input. Yellow arrows highlight the anatomical regions affected by the simulated deformation.}
    \vspace*{2ex}
    \label{fig:teaser}
\end{figure*}

In summary, our contributions are as follows:

\begin{itemize}
    \item \textcolor{black}{We propose a semantic-label-driven Stage~I approach for multi-organ abdominal US synthesis that generates simplified US images directly from anatomical masks at inference time, approximating the physics-based ray-casting outputs used during training (Fig.~\ref{fig:teaser});}

    \item We formulate Stage~I as a paired label-to-physics-based image translation problem, where generative models learn to approximate CT-derived ray-casting outputs after being trained with physics-based supervision. This allows the trained model to avoid executing ray-casting during inference;

    \item We compare two major generative paradigms, GAN-based and diffusion-based models, for Stage~I synthesis in terms of pixel-level accuracy, perceptual realism, anatomical preservation, and generation time. This comparison provides practical insight into the trade-off between visual realism, anatomical fidelity, and simulation efficiency;

    \item \textcolor{black}{We evaluate whether the outputs generated by the new Stage~I models remain compatible with the previously established SG-CycleGAN refinement stage~\cite{vitaleimproving}, using it as a fixed downstream module. The proposed modular framework allows paired and unpaired image-to-image translation models to be coupled or decoupled, enabling progressive control over image structure and appearance;}

    \item We illustrate how semantic label manipulation can be used to generate controlled \textcolor{black}{proof-of-concept} simulations, including anatomical deformations and synthetic pathological findings such as liver tumors, kidney stones, and gallbladder polyps.
       
\end{itemize}

\section{Materials and Methods}
\label{sec:Materials_and_Methods}

\begin{figure}[t]
    \centering
    \includegraphics[width=1\textwidth]{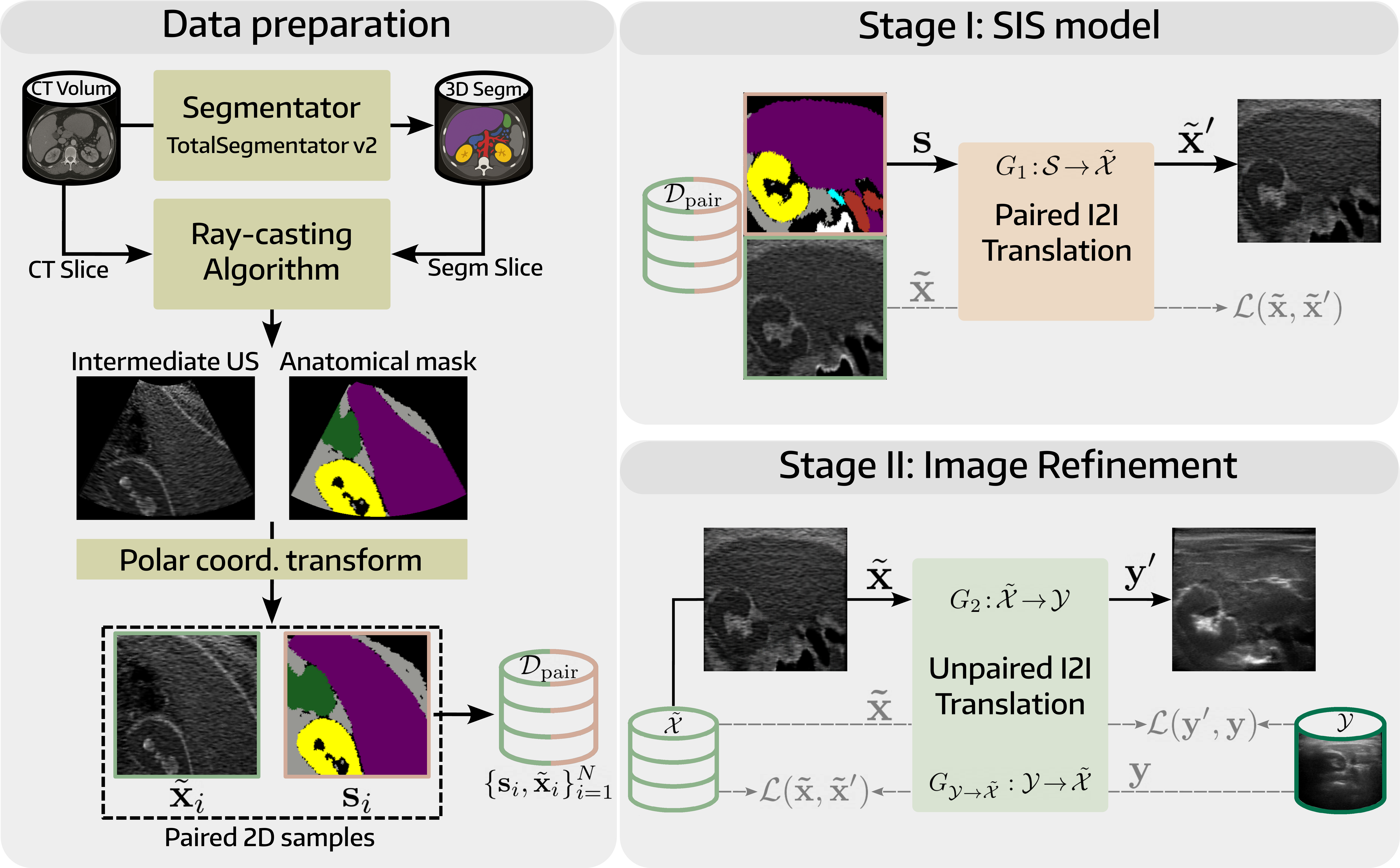}
    \caption{Overview of the experimental framework. The left block shows data preparation from CT-derived segmentations and ray-casting simulations. \textcolor{black}{Stage~I is evaluated using two alternative implementations, Pix2Pix and SDM, while Stage~II uses the previously established SG-CycleGAN refinement strategy.}}
    \vspace*{2ex}
    \label{fig:schematic}
\end{figure}

A schematic representation of the proposed approach is depicted in Fig.~\ref{fig:schematic}. Simplified physics-based 2-D US images are first generated from pairs of 3-D CT scans and their associated segmentations by means of a ray-casting simulator (Section~\ref{methods:data-prepro}). These images $\tilde{\mathbf{x}}_{i}\!\in\!\tilde{\mathcal{X}}$—with $\tilde{\mathcal{X}}$ denoting the domain of all synthetic scans—are naturally paired with anatomical masks $\mathbf{s}_{i}\!\in\!\mathcal{S}$—$\mathcal{S}$ being the set of all possible segmentation masks. Hence a paired data set $\mathcal{D}_{\text{pair}}=\{(\mathbf{s}_{i},\tilde{\mathbf{x}}_{i})\}_{i=1}^{N}$ can be used to train a conditional generator $G_{1}\!:\!\mathcal{S}\!\rightarrow\!\tilde{\mathcal{X}}$ that maps a mask $\mathbf{s}\in\mathcal{S}$ to a synthetic US slice $\tilde{\mathbf{x}}\in\tilde{\mathcal{X}}$ (Section~\ref{subsec:methods-paired}). This model captures tissue-specific echogenicities and common US artefacts such as reflections and occlusions while bypassing any CT input at inference. However, the resulting images remain visually unrealistic when compared with clinical scans (Fig.~\ref{fig:arqui}). To mitigate this, the overall mapping is expressed as the composition $G = G_{2}\!\circ\!G_{1}$, where $G_{2}\!:\!\tilde{\mathcal{X}}\!\rightarrow\!\mathcal{Y}$ refines a synthetic image $\tilde{\mathbf{x}}$ into a realistic US image $\mathbf{y}\!\in\!\mathcal{Y}$; $G_{1}$ and $G_{2}$ are trained independently. Because paired ground truth for this refinement step is scarce, $G_{2}$ is learned with an unpaired image-to-image translation scheme (Section~\ref{subsec:methods-unpaired}). The second stage enhances speckle texture, scattering effects, and introduces artifacts absent from the intermediate domain, yielding images that better match real US appearance (Fig.~\ref{fig:arqui}). At inference, the complete pipeline enables the simulation of realistic US sequences from any input segmentation—allowing breathing motion, synthetic lesions, and operator-induced deformations such as abdominal pressure to be modeled (Fig.~\ref{fig:teaser}C).

\subsection{Data preparation}
\label{methods:data-prepro}

\subsubsection{Materials}
Thirty-nine abdominal CT scans were collected from two public sources: the VISCERAL Anatomy3 Challenge dataset (healthy subjects)~\cite{Jimenez2016} and the IRCAD Liver Segmentation dataset~\footnote{\url{https://www.ircad.fr/research/data-sets/liver-segmentation-3d-ircadb-01/}} (healthy and diseased subjects). This last cohort was 54 \% male, and included 41 \% of cases with liver tumors.

In addition, we used the same 617 B-mode US images from \cite{vitaleimproving} as $\mathcal{Y}$. These scans were acquired from 11 healthy volunteers (60 \% male, age $27\pm3$ years) during routine abdominal examinations with a SonoSite M-Turbo v1.3 scanner (FUJIFILM, Bothell, USA), and exported at a resolution of $640\times480$ px.

\subsubsection{CT segmentation}
\label{methods:segmentation}
Each CT volume was automatically segmented with \texttt{TOTALSEGMENTATOR} v2~\cite{wasserthal2023totalsegmentator}, yielding masks for 16 anatomical structures (bones, arteries, pancreas, gallbladder, kidneys, adrenal glands, liver, spleen, veins, colon, duodenum, small bowel, stomach, visceral fat, liver tumors, background). For IRCAD cases we used the manual annotations of liver tumors released with that dataset~\cite{soler20103d}. The same anatomical label convention and color mapping were used throughout all model inputs and figures. The complete color-code reference is provided in Supplementary Fig.~1.

\subsubsection{Generation of simplified US images with a physics-based simulator}
Simplified US images ($\tilde{\mathcal{X}}$) were generated from the segmented CT volumes using the ray-casting approach described by Rubi \textit{et al.}~\cite{rubi2017comparison}. Our simulator uses a virtual probe to define the acquisition position, orientation, and direction and extract a 2D slice from the 3D volume. This image is used as input for an algorithm that simulates the propagation of the US wave through the tissue, capturing reflections, transmissions, and scattering effects at organ interfaces. The combination of these interactions produces the final synthetic US image. As image intensities depend directly on the CT-derived Hounsfield Unit (HU) values, a correction was applied to the transmission window following the method used in~\cite{vitaleimproving}, which adjusts tissue specific transmissions using the liver's value as a reference anchor. 
This simulation strategy was selected due to its ability to produce anatomically plausible US appearances while remaining compatible with the unpaired refinement architectures (Stage II). Full implementation details and echogenicity correction coefficients are provided in~\cite{vitaleimproving}.

\subsubsection{Final dataset assembly}
\label{methods:dataset}
The pipeline produced $N = 1036$ paired \textcolor{black}{2D} samples $\{(\mathbf{s}_{i}, \tilde{\mathbf{x}}_{i})\}_{i=1}^{N} \subset \mathcal{D}_{\text{pair}}$, where each sample consists of a semantic mask and its corresponding physics-based synthetic US image. These samples were generated from 39 CT volumes and split at the patient/volume level, ensuring that all 2D samples from a given CT volume belonged to a single partition. The resulting partitions contained 711 training samples, 80 validation samples, and 245 test samples, corresponding to 27, 4, and 8 CT volumes, respectively. The validation set was used for model selection and hyperparameter adjustment, whereas the held-out test set was used for the final quantitative evaluation of $G_{1}$ (Section~\ref{subsec:methods-paired}). Furthermore, the synthetic domain $\tilde{\mathcal{X}}$ obtained from $\mathcal{D}_{\text{pair}}$ was combined with the real-image domain $\mathcal{Y}$ to train $G_{2}$ (Section~\ref{subsec:methods-unpaired}).
All samples were represented in polar coordinates during model training and inference. Images were grayscale and intensity values were represented in the range 0--255. Semantic masks were encoded as label maps, where each class color was associated with a discrete anatomical label. Pix2Pix was trained to generate $256\times256$ images directly. SDM generated $128\times128$ images to reduce inference cost, and outputs were resized to $256\times256$ for quantitative comparison with Pix2Pix and the physical-model reference images.

\subsection{Paired segmentation to physics-based image translation}
\label{subsec:methods-paired}

\begin{figure}[t]
    \centering
    \includegraphics[width=1\textwidth]{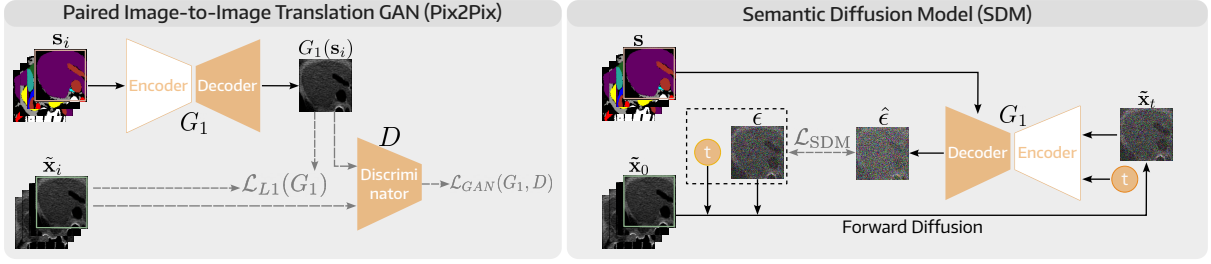}
    \caption{Training schemes for the Stage~I paired models. Left: Pix2Pix is trained with adversarial and L1 losses in a paired image-to-image setting. Right: SDM performs iterative denoising guided by anatomical labels through spatially adaptive normalization.}
    \vspace*{2ex}
    \label{fig:arqui}
\end{figure}

Module $G_{1}$ in Fig.~\ref{fig:schematic} is a paired image-to-image translation block trained with $\mathcal{D}_{\text{pair}}$ that maps a segmentation mask $\mathbf{s}_{i}$ to a synthetic US image $\tilde{\mathbf{x}}_{i}\in\tilde{\mathcal{X}}$. A GAN-based solution, specifically Pix2Pix~\cite{isola2017image}, and a diffusion-based alternative, SDM~\cite{wang2022semantic}, were selected to compare two complementary generative paradigms for paired label-to-image translation. Pix2Pix provides a strong feed-forward conditional GAN baseline, widely used in medical image-to-image translation and suitable for fast generation and high-frequency texture synthesis. In contrast, SDM represents a diffusion-based alternative with semantic conditioning and iterative denoising, which may better preserve anatomical layout. This design allows us to evaluate the practical trade-off between perceptual realism, anatomical fidelity, and computational cost in Stage~I. Other generative families, including VAEs, flow-based models, and ControlNet-like conditional diffusion models, were not included in this first evaluation. VAEs and flows introduce different likelihood-based modeling assumptions, whereas ControlNet variants commonly rely on pretrained diffusion backbones and domain adaptation. Figure~\ref{fig:arqui} sketches the training set-ups for both cases, and each configuration is detailed in the sequel.

\subsubsection{Pix2Pix}
Pix2Pix~\cite{isola2017image} was used as the GAN-based implementation of $G_{1}$ because it is a widely adopted conditional image-to-image translation model, including in US synthesis~\cite{Pigeau,zhang2020deep}.  As shown in Fig.~\ref{fig:arqui}, the anatomical label map $\mathbf{s}_{i}$ conditions the generation of the corresponding physics-based US image $\tilde{\mathbf{x}}_{i}$. The model was trained with the standard adversarial and reconstruction losses:
\begin{equation}
\label{eq:pix2pix_final}
\mathcal{L}_{\text{Pix2Pix}}(G_{1},D)=
\mathcal{L}_{\text{GAN}}(G_{1},D)
+
\lambda_{\text{L1}}
\mathcal{L}_{\text{L1}}(G_{1}),
\end{equation}
where $\mathcal{L}_{\text{GAN}}(G_{1},D)$ encourages realistic outputs  through the discriminator $D$, while the reconstruction term $\mathcal{L}_{\text{L1}}(G_{1})=
\mathbb{E}_{(\mathbf{s},\tilde{\mathbf{x}})\sim\mathcal{D}_{\text{pair}}}
\bigl[\,
\lVert \tilde{\mathbf{x}}-G_{1}(\mathbf{s})\rVert_{1}
\bigr]
$
penalizes pixel-wise differences between generated and reference images. The hyper-parameter $\lambda_{\text{L1}}$ balances both terms.
After training, only $G_{1}$ is used at inference to produce synthetic US images from unseen anatomical label maps—without requiring CT volumes or ray-casting simulation.

We implemented $G_{1}$ as a U-Net with seven down-sampling/up-sampling blocks with skip connections. The first four encoder blocks progressively halve the spatial resolution via strided $4{\times}4$ convolutions (stride = 2) while increasing feature maps to 64, 128, 256, and 512; each block applies batch normalization and Leaky-ReLU with a 0.2 negative slope. The last three encoder blocks employ 512-channel convolutions. Each decoder block performs nearest-neighbor up-sampling followed by a $3{\times}3$ convolution (stride = 1), batch normalization, and ReLU; a final $\tanh$ activation generates the output image. The discriminator $D$ follows the $70{\times}70$ PatchGAN design of \cite{isola2017image}, mirroring the encoder’s down-sampling path. Training used a batch size of 64 for 2\,000 epochs with Adam optimization ($\mathrm{lr}=2\times10^{-4}$, $\beta_1=0.9$, $\beta_2=0.999$). The L1 weight was set to $\lambda_{\text{L1}}=100$, following the original Pix2Pix configuration~\cite{isola2017image}.
Model checkpoints were evaluated on the validation set. The final Pix2Pix checkpoint was selected using LPIPS as the primary validation metric, complemented by visual inspection of anatomical consistency and image quality.

\subsubsection{Semantic Diffusion Model (SDM)}

We adopted the SDM of Wang \textit{et al.}~\cite{wang2022semantic} as the diffusion-based implementation of $G_{1}$. SDM follows the Denoising Diffusion Probabilistic Model (DDPM) framework~\cite{ho2020denoising}, where a forward process corrupts $\tilde{\mathbf{x}}$ with Gaussian noise and a learnable reverse process denoises it step by step under the guidance of the anatomical label map $\mathbf{s}$. A cosine noise schedule~\cite{nichol2021improved} was used to stabilize training and improve quality. 
As shown in Fig.~\ref{fig:arqui}, the SDM training loop corrupts the physics-based US image with Gaussian noise and trains the denoising network $G_{1}$ to predict the added noise at each time step. The noisy image is processed by the encoder, while the label map $\mathbf{s}$ is injected into the decoder through multi-layer spatially adaptive normalization (SPADE) blocks~\cite{park2019semantic}. Training minimizes:
\begin{equation}
\label{eq:sdm_final}
\mathcal{L}_{\text{SDM}}
=\!
\mathbb{E}_{t,\;(\mathbf{s},\tilde{\mathbf{x}})\sim\mathcal{D}_{\text{pair}}}
\bigl[
\|\,\hat{\epsilon}_\theta(\tilde{\mathbf{x}}_t,\mathbf{s},t)-\epsilon\|_2^{\,2}
\bigr]
\;+\;
\lambda_{\text{KL}}
\,\mathrm{KL}\!\bigl(q_t\;\|\;p_\theta\bigr),
\end{equation}
where the first term is the conditional noise-prediction loss. Here, $\epsilon!\sim!\mathcal{N}(0,\mathbf{I})$ is the known Gaussian noise added to $\tilde{\mathbf{x}}$ at timestep $t$, and $\hat{\epsilon}{\theta}$ is the predicted noise. The second term is a KL regularization term weighted by $\lambda{\text{KL}}$.

During inference, sampling starts from pure noise and iteratively denoises under the guidance of $\mathbf{s}$ to obtain the final image $G_{1}(\mathbf{s})$, enabling controllable US synthesis without executing CT-derived ray-casting simulations.

The backbone is an ADM-style U-Net with 64 base channels, channel multipliers $(1,1,2,3,4)$, and two residual blocks per resolution level. Self-attention was applied at $32{\times}32$, $16{\times}16$, and $8{\times}8$ resolutions. The encoder uses residual blocks conditioned on the diffusion timestep, while the bottleneck and decoder use SDMResBlocks with SPADE-based normalization to inject the semantic condition. The semantic label map is encoded as a 16-channel one-hot tensor and used to modulate the decoder features. The model predicts the added noise $\epsilon$, and was trained with a cosine noise schedule, mixed precision, EMA ($0.9999$), Adam optimization ($\mathrm{lr}=1\times10^{-4}$), batch size 12, and $\lambda_{\text{KL}}=0.001$. The final SDM checkpoint was selected from validation performance using LPIPS as the primary metric, complemented by visual inspection of anatomical consistency and image quality. We evaluated 250, 500, 750, and 1000 denoising steps, as reported in the Supplementary Material, and used the EMA checkpoint at 460000 training steps for the final experiments.

\subsection{Unpaired physics-based to realistic image translation}
\label{subsec:methods-unpaired}

To enhance the visual realism of Stage~I outputs, \textcolor{black}{we retrained a segmentation-guided CycleGAN (SG-CycleGAN) using the architecture and training strategy introduced by Vitale \textit{et al.}~\cite{vitaleimproving}, with the datasets described in Section~\ref{methods:dataset}}. This model extends CycleGAN~\cite{zhu2017unpaired} with a segmentation-consistency term to encourage preservation of organ geometry during unpaired translation.

Let $\tilde{\mathcal{X}}$ be the domain of synthetic images produced in Stage I and let $\mathcal{Y}$ denote the domain of real US scans. The network learns two mappings:  
\[
G_{2}=G_{\tilde{\mathcal{X}}\rightarrow\mathcal{Y}} : \tilde{\mathcal{X}}\!\rightarrow\!\mathcal{Y},
\qquad
G_{\mathcal{Y}\rightarrow\tilde{\mathcal{X}}} : \mathcal{Y}\!\rightarrow\!\tilde{\mathcal{X}},
\]
together with discriminators $D_{\tilde{\mathcal{X}}}$ and $D_{\mathcal{Y}}$. Training minimizes:
\begin{equation}
\label{eq:sgcyclegan_final}
\mathcal{L}_{\text{SG-CycleGAN}}
=
\mathcal{L}_{\text{GAN}}
+
\lambda_{\text{cyc}}\mathcal{L}_{\text{cyc}}
+
\lambda_{\text{idt}}\mathcal{L}_{\text{idt}}
+
\lambda_{\text{sg}}\mathcal{L}_{\text{sg}},
\end{equation}
where $\lambda_{\text{cyc}}$, $\lambda_{\text{idt}}$, and $\lambda_{\text{sg}}$ weight the respective terms.

Unpaired samples $\tilde{\mathbf{x}}\!\sim\!\tilde{\mathcal{X}}$ and $\mathbf{y}\!\sim\!\mathcal{Y}$ feed the generators, while a pre-trained segmentation network $S$ supplies masks for the segmentation-guided loss $\mathcal{L}_{\text{sg}}$, encouraging structural alignment.  
At inference, only $G_{2}$ is used: a Stage I image $\tilde{\mathbf{x}}$ is transformed into a realistic scan $G_{2}(\tilde{\mathbf{x}})$ without requiring paired real US data.

The SG-CycleGAN model was trained from scratch following the architecture and training regimen used in~\cite{vitaleimproving}. The loss weights were set to $\lambda_{\text{cyc}}=10$, $\lambda_{\text{idt}}=0.5$, and $\lambda_{\text{sg}}=0.5$, based on validation-set performance using FID and visual inspection of anatomical consistency. The model was trained for 200 epochs with Adam optimization, an initial learning rate of $2\times10^{-4}$, and batch size 4. After 100 epochs, the learning rate was linearly decayed to zero.  The segmentation network $S$ was trained offline to predict 16 abdominal structures (Section~\ref{methods:data-prepro}) using multi-class cross-entropy. It was optimized with Adam, an initial learning rate of $1\times10^{-4}$, batch size 16, and 150 training epochs. The learning rate was reduced by a factor of 0.5 after 20 validation-plateau epochs, measured by the average Dice coefficient. Importantly, the segmenter $S$ was trained independently of Pix2Pix and SDM using simplified physics-based US images with CT-derived masks as supervision, and was not trained on images generated by either Stage~I model.

\subsection{Inference}

Given an anatomical label map $\mathbf{s}\in\mathcal{S}$, the pipeline is used at inference time to produce a realistic US image in two steps: first, $\tilde{\mathbf{x}} = G_{1}(\mathbf{s})$, and then $\mathbf{y} = G_{2}(\tilde{\mathbf{x}})$. As both generators work in polar coordinates, the resulting output image $\mathbf{y}$ is finally resampled to Cartesian space to be presented in a standard manner.

The input $\mathbf{s}$ is assumed to be an anatomically plausible semantic label map. It must be represented in polar coordinates at $256\times256$ resolution and encoded using the same discrete anatomical-label convention and color mapping used during training. Such masks may come from reusable anatomical templates, atlas-based models, manually edited cases, procedurally generated anatomies, or deformed versions of existing labeled anatomies.
\textcolor{black}{To generate controlled dynamic or pathological scenarios, a user-defined transformation $t$ can be applied to the input label map before image synthesis, yielding $\tilde{\mathbf{s}} = t(\mathbf{s})$. We simulated breathing-like motion by deforming the 3D semantic label volumes over eight temporal states using a simplified parametric displacement field. The deformation amplitude varied sinusoidally over time and was spatially modulated by a Gaussian profile to produce smooth, non-rigid motion concentrated toward the anterior side of the volume. Each anatomical label was warped independently using nearest-neighbor interpolation to preserve discrete label values, while the bone class was kept fixed as a rigid reference. This procedure provides a cyclic approximation of respiratory motion and should be interpreted as a simplified deformation model rather than a biomechanical simulation of abdominal breathing.
Pathological scenarios were generated by manually editing the 2D semantic label maps in 3D Slicer. Pathological regions were introduced by modifying existing anatomical labels or adding new labeled regions while preserving the same color-coded semantic convention used during training. The edits were manually designed by the authors to create anatomically plausible proof-of-concept examples, rather than being randomly sampled or derived from a formal clinical protocol.} In the experiments reported here, the starting masks were held-out anatomical label maps from the test set, which were used only as semantic inputs to the pipeline. The corresponding physics-based simulations were used only as references for Stage~I evaluation. For pathological cases, examples of liver tumors, abdominal aortic aneurysms, kidney stones, and gallbladder polyps were generated as described in Section~\ref{sec:Patho}.

\subsection{Evaluation metrics and statistical analysis}
\label{subsec:stats}

Evaluating generative image models is inherently challenging because no single metric can fully capture image fidelity, anatomical correctness, perceptual realism, and distributional similarity. This is particularly relevant in US simulation, where images contain stochastic speckle patterns and acoustic artifacts that may differ locally even when the underlying anatomy and clinically relevant appearance are correctly preserved. \textcolor{black}{We therefore followed common practice in the US simulation literature and combined complementary metrics designed to assess different properties of the generated images~\cite{Zhang2021, Tomar2021, liang2022sketch, katakis2023generation,  peng2025denoising, du2026synthetic, dominguez2024diffusion, chen2023anatomy}}.

For Stage~I, simulated images were compared with their paired physics-based reference images $\tilde{\mathbf{x}}$. \textcolor{black}{The evaluation metrics were grouped into two complementary categories: (i) morphological and anatomical metrics, including Mean Absolute Error (MAE), Mean Squared Error (MSE), Peak Signal-to-Noise Ratio (PSNR), Structural Similarity Index (SSIM), and mean Intersection over Union (mIoU), which assess reconstruction fidelity and preservation of spatial and anatomical structure; and (ii) perceptual and distributional metrics, including Learned Perceptual Image Patch Similarity (LPIPS), Fréchet Inception Distance (FID), and Kernel Inception Distance (KID), which assess feature-level similarity and agreement between image distributions.}
Intensity fidelity was quantified using MAE, MSE, and PSNR. These metrics measure pixel-wise differences between generated and reference images, with MAE and MSE directly quantifying reconstruction error and PSNR expressing this error relative to the available intensity range~\cite{Zhang2021,katakis2023generation,peng2025denoising}. Structural similarity was evaluated using SSIM, which compares local luminance, contrast, and structural patterns~\cite{Zhang2021,Tomar2021,chen2023anatomy,peng2025denoising}.
Because pixel-wise similarity does not necessarily imply preservation of anatomical content, we also evaluated anatomical consistency using mIoU across 16 abdominal structures. The same segmentation model $S$ was applied to each physics-based reference image and its corresponding simulated image, and the resulting predictions were compared. The mIoU therefore measures whether anatomical structures detectable in the reference image remain consistently represented after image synthesis. As this evaluation depends on the robustness of $S$ on generated images, mIoU was interpreted as a relative measure of segmentation consistency rather than as an independent estimate of anatomical accuracy~\cite{zhang2023unpaired}. 
Perceptual similarity was assessed using LPIPS, which measures discrepancies between deep feature representations and is less dependent on exact pixel correspondence~\cite{dominguez2024diffusion,peng2025denoising}. Distribution-level similarity was assessed using FID and KID. Unlike the preceding image-level metrics, FID and KID compare feature distributions extracted from the complete sets of reference and generated images rather than individual image pairs. FID is widely used to evaluate generative models, whereas KID provides an unbiased estimator that can be more appropriate for relatively small datasets. Although both metrics rely on an Inception~v3 network pretrained on ImageNet and are therefore not specific to US, they have been used in previous US synthesis studies~\cite{liang2022weakly,liang2022sketch,Tomar2021,Zhang2021}. Following previous work~\cite{Tomar2021}, features were extracted from the intermediate 768-dimensional layer to reduce reliance on the final, ImageNet-specific representation. FID and KID were computed using the TorchFidelity package~\cite{obukhov2020torchfidelity}.

\textcolor{black}{Stage~II was evaluated here primarily as a downstream compatibility test, to determine whether the images generated by the proposed Stage~I models produce refined outputs comparable to those obtained when the same refinement module is applied to physics-based ray-casting images.}
Stage~II could not be evaluated using paired reconstruction metrics because no corresponding real US image was available for each refined image. Its evaluation therefore focused on similarity to the real US domain. Global distributional similarity was quantified using FID and KID, while organ-specific echogenicity was evaluated using $\chi^2$ distances between intensity histograms extracted from selected anatomical regions. Lower histogram distances indicate greater agreement between the intensity distributions of generated and real tissues. Because these quantitative measures do not directly capture ultrasound-specific artifact correctness or clinical plausibility, they were interpreted together with anatomy-aware consistency measures and qualitative inspection of relevant acoustic phenomena, including acoustic shadows, anechoic regions, and hyperechoic responses.

For the statistical analysis of Stage~I, MAE, MSE, PSNR, SSIM, LPIPS, and mIoU were computed independently for each test image. Pix2Pix and SDM were therefore compared using paired vectors containing one observation per image. Differences between methods were assessed using one-tailed Wilcoxon signed-rank tests, with statistical significance defined as $p<0.05$. FID and KID were treated as dataset-level metrics because each value summarizes the complete reference and generated image distributions. Consequently, they were not subjected to image-level hypothesis testing. Instead, their uncertainty was estimated using 95\% bootstrap confidence intervals based on 1,000 resamples. The same bootstrap procedure was used for FID and KID in the Stage~I and Stage~II evaluations.


\section{Results}
\label{sec:Results}

We evaluated each stage of the proposed pipeline both quantitatively and qualitatively. For Stage~I, we compared the two generative models described in Section~\ref{subsec:methods-paired}, namely Pix2Pix and SDM. SDM was run with $750$ inference steps, which we found to be a cost-effective trade-off between generation time and image quality (see the supplementary material for details). We also assessed the integration with Stage~II by using the refinement model (Section~\ref{subsec:methods-unpaired}) to translate outputs from each Stage~I method. All models were implemented in PyTorch~2.0.1. Training and inference experiments were performed on a desktop workstation running Ubuntu~18.04.6, equipped with an Intel Core i7-10700K CPU, 16~GB of RAM, and a single NVIDIA Quadro P6000 GPU with 24~GB of VRAM.

\subsection{Quantitative Evaluation}

\subsubsection{Stage I: Simplified US Simulation}

\begin{table}[ht]
\caption{Quantitative comparison of Pix2Pix and SDM. Paired image-level metrics are reported as mean $\pm$ standard deviation over the test set. FID and KID are dataset-level metrics and are reported with bootstrap-derived 95\% confidence intervals. Bold values indicate the best point estimate. Asterisks (*) indicate statistically significant differences between Pix2Pix and SDM for paired image-level metrics ($p < 0.05$).}
\label{table:Quantitative}

\centering
\scriptsize
\setlength{\tabcolsep}{5.2pt}
\renewcommand{\arraystretch}{1.35}

\begin{tabular}{
>{\centering\arraybackslash}m{1.25cm} 
>{\centering\arraybackslash}m{1.20cm} 
>{\centering\arraybackslash}m{1.25cm} 
>{\centering\arraybackslash}m{1.20cm} 
>{\centering\arraybackslash}m{1.15cm} 
>{\centering\arraybackslash}m{1.15cm} 
@{\hspace{5pt}}
>{\centering\arraybackslash}m{1.20cm} 
>{\centering\arraybackslash}m{1.50cm} 
>{\centering\arraybackslash}m{1.50cm} 
}
\toprule

\textbf{Model} &
\multicolumn{5}{c}{\textbf{Morphological metrics}} &
\multicolumn{3}{c}{\textbf{Perceptual metrics}} \\

\cmidrule(lr){2-6}
\cmidrule(lr){7-9}

&
\makecell[c]{MAE\\($\downarrow$)} &
\makecell[c]{MSE\\($\downarrow$)} &
\makecell[c]{PSNR\\($\uparrow$)} &
\makecell[c]{SSIM\\($\uparrow$)} &
\makecell[c]{mIoU\\($\uparrow$)} &
\makecell[c]{LPIPS\\($\downarrow$)} &
\makecell[c]{FID\\($\downarrow$)} &
\makecell[c]{KID\\($\times10^{-3}$)} \\

\midrule

Pix2Pix &
\makecell[c]{21.65\\$\pm$ 2.95} &
\makecell[c]{884.72\\$\pm$ 246.11} &
\makecell[c]{18.81\\$\pm$ 1.12} &
\makecell[c]{0.24\\$\pm$ 0.04} &
\makecell[c]{0.29\\$\pm$ 0.11} &
\makecell[c]{\textbf{0.17}*\\$\pm$ 0.04} &
\makecell[c]{\textbf{0.32}\\{\textbf{[0.30--0.35]}}} &
\makecell[c]{\textbf{0.25}\\{\textbf{[0.21--0.29]}}} \\

\midrule

SDM &
\makecell[c]{\textbf{19.85}*\\$\pm$ 2.74} &
\makecell[c]{\textbf{769.18}*\\$\pm$ 219.45} &
\makecell[c]{\textbf{19.43}*\\$\pm$ 1.16} &
\makecell[c]{\textbf{0.28}*\\$\pm$ 0.04} &
\makecell[c]{\textbf{0.43}*\\$\pm$ 0.15} &
\makecell[c]{0.19\\$\pm$ 0.03} &
\makecell[c]{0.37\\{[0.34--0.40]}} &
\makecell[c]{0.48\\{[0.41--0.56]}} \\

\bottomrule

\end{tabular}
\end{table}

\textcolor{black}{We quantitatively evaluated both Stage~I models using two complementary groups of metrics: morphological metrics, which assess reconstruction fidelity and anatomical preservation, and perceptual metrics, which assess feature-level and distributional similarity.} The results are summarized in Table~\ref{table:Quantitative}, where asterisks (*) indicate statistically significant differences for paired image-level metrics only, as described in Section~\ref{subsec:stats}. FID and KID are reported with bootstrap-derived 95\% confidence intervals.

\textcolor{black}{Regarding the morphological metrics, SDM consistently outperformed Pix2Pix. It achieved lower MAE (19.85 vs. 21.65) and MSE (769.18 vs. 884.72), together with higher PSNR (19.43 vs. 18.81) and SSIM (0.28 vs. 0.24). These results indicate a more faithful reconstruction of the physics-based reference images at both pixel and structural levels. SDM also achieved a substantially higher mIoU than Pix2Pix (0.43 vs. 0.29), indicating better preservation of anatomical structures represented in the physics-based reference images.}

In contrast, Pix2Pix obtained better perceptual and distributional scores. It achieved a significantly lower LPIPS than SDM (0.17 vs. 0.19). Pix2Pix also yielded lower FID (0.32 vs. 0.37) and KID (0.25 vs. 0.48) point estimates. The KID confidence intervals did not overlap, indicating a clear distribution-level difference between the two models. In contrast, the FID confidence intervals partially overlapped and should therefore be interpreted as a descriptive difference rather than as evidence of statistical significance. \textcolor{black}{Overall, Pix2Pix showed marginally better perceptual and distribution-level similarity, whereas SDM demonstrated substantially stronger morphological and anatomical consistency.}

\subsubsection{Stage II: Image Refinement}

To evaluate how well the SIS models integrate with the refinement stage, we used the SG-CycleGAN refiner to enhance images generated by each Stage~I method (Pix2Pix and SDM), following the protocol of Vitale \textit{et al.}~\cite{vitaleimproving}. The same SG-CycleGAN was applied to three inputs (images from the physical model, Pix2Pix, and SDM) allowing us to assess both the compatibility of SIS outputs with the refiner and the incremental contribution of Stage~II.

Table~\ref{tab:realism} reports FID, KID, and $\chi^{2}$ distances for simulated images before ($\tilde{\mathbf{x}}$) and after ($\tilde{\mathbf{y}}$) Stage~II refinement, using real US scans ($\mathbf{y}$) as the reference domain. The refinement stage substantially improved distribution-level realism for all inputs. For Pix2Pix-derived images, FID decreased from 1.66 to 0.39 and KID from 4.92 to 0.30. For SDM-derived images, FID decreased from 2.09 to 0.53 and KID from 6.28 to 0.84. A similar improvement was observed for the physical-model images, where FID decreased from 1.81 to 0.38 and KID from 5.19 to 0.29. These results quantify the added value of Stage~II in terms of distributional alignment with real US images.
When considering only the refined outputs, the physical-model input yielded the best FID and KID values, as expected because the SG-CycleGAN refiner was trained using physical-model images as the synthetic source domain. Among the SIS inputs, Pix2Pix-refined images were closer to the real US distribution than SDM-refined images, with lower FID (0.39 vs. 0.53) and KID (0.30 vs. 0.84). SDM-refined images showed higher FID and KID than the refined physical-model outputs based on non-overlapping 95\% confidence intervals, whereas Pix2Pix-refined images showed overlapping intervals with the physical-model outputs. This suggests that Pix2Pix outputs are more compatible with the texture distribution learned by the refiner, while SDM outputs remain smoother and farther from the physical-model source domain.
Echogenicity consistency was assessed as in~\cite{vitaleimproving} by computing $\chi^{2}$ distances between intensity histograms in three representative anatomical regions (liver, kidney, and gallbladder) for each image set versus real scans. Compared with the large improvements observed for FID and KID, the effect of Stage~II on organ-level echogenicity was smaller and more region-dependent. Refinement improved the liver $\chi^{2}$ distance for Pix2Pix (0.24 to 0.18) and SDM (0.29 to 0.19), and also reduced the kidney $\chi^{2}$ distance for both Pix2Pix (0.29 to 0.26) and SDM (0.26 to 0.23). In the gallbladder, the values were already close to the real-US reference range before refinement and changed only marginally after Stage~II. These results suggest that SG-CycleGAN mainly improves distribution-level realism and texture appearance, while preserving or moderately adjusting the organ-level echogenicity patterns inherited from Stage~I.
When analyzing only the refined outputs, all three refined sets were statistically indistinguishable from real US in the gallbladder region in terms of intensity distribution ($p > 0.5$). In the liver, Pix2Pix-refined and physical-model-refined images preserved realistic intensity distributions (no significant difference from real, $p > 0.05$), whereas SDM-refined images showed a significant divergence from real data ($p < 0.01$). Notably, the SDM-refined liver intensity distribution matched the real mean and 95\% interval most closely, but showed higher variance, leading to more dispersed intensities, as evidenced by the boxplot in Supplementary Fig.~3. In the kidney, both Pix2Pix-refined and SDM-refined outputs were closer to the real intensity distribution than the refined physical-model images, yielding smaller $\chi^{2}$ distances on average. SDM-refined images had the lowest $\chi^{2}$ divergence in the kidney region, although the difference from real scans remained statistically significant ($p \approx 0$ in all cases).

\begin{table}[t]
\caption{Quantitative comparison of simulated images before and after Stage~II refinement against real US scans. FID and KID measure distribution-level similarity to the real US domain, while $\chi^2$ distances compare organ-level intensity histograms with real US scans. Lower scores ($\downarrow$) indicate better agreement. In the $\chi^2$ columns, asterisks (*) mark $p<0.01$ versus real US, and daggers (\dag) denote $p>0.05$ versus real US. KID values are reported as $\times 10^{-3}$.}
\label{tab:realism}
\centering
\footnotesize
\setlength{\tabcolsep}{2.2pt}
\renewcommand{\arraystretch}{1.25}
\begin{tabular}{
>{\centering\arraybackslash}p{1.9cm}
>{\centering\arraybackslash}p{0.9cm}
>{\centering\arraybackslash}p{2.1cm}
>{\centering\arraybackslash}p{2.0cm}
>{\centering\arraybackslash}p{2.1cm}
>{\centering\arraybackslash}p{2.1cm}
>{\centering\arraybackslash}p{2.0cm}
}
\toprule
Image set
& Stage~II
& FID $\downarrow$ [95\% CI]
& KID $\downarrow$ [95\% CI]
& Liver $\chi^2$ [95\% CI]
& Kidney $\chi^2$ [95\% CI]
& Gallbladder $\chi^2$ [95\% CI] \\
\midrule

Physical model & No
& 1.81 [1.76--1.86]
& 5.19 [5.00--5.39]
& 0.24 [0.03--0.53]
& 0.26 [0.04--0.49]
& 0.19 [0.00--0.61] \\

Pix2Pix & No
& 1.66 [1.61--1.71]
& 4.92 [4.74--5.11]
& 0.24 [0.04--0.52]
& 0.29 [0.03--0.52]
& 0.25 [0.02--0.64] \\

SDM & No
& 2.09 [2.03--2.16]
& 6.28 [6.03--6.54]
& 0.29 [0.09--0.50]
& 0.26 [0.06--0.49]
& 0.22 [0.00--0.54] \\

\midrule

Physical model & Yes
& 0.383 [0.36--0.40]
& 0.29 [0.25--0.34]
& 0.18 [0.04--0.40]\dag
& 0.28 [0.04--0.58]
& 0.24 [0.00--0.50]\dag \\

Pix2Pix & Yes
& 0.385 [0.36--0.40]
& 0.30 [0.25--0.34]
& 0.18 [0.04--0.39]\dag
& 0.26 [0.04--0.59]*
& 0.25 [0.03--0.54]\dag \\

SDM & Yes
& 0.531 [0.49--0.56]
& 0.84 [0.73--0.95]
& 0.19 [0.04--0.43]*
& 0.23 [0.03--0.52]*
& 0.25 [0.04--0.63]\dag \\

\midrule
Real US & --
& --
& --
& 0.20 [0.01--0.51]
& 0.18 [0.00--0.45]
& 0.24 [0.00--0.48] \\
\bottomrule
\end{tabular}
\end{table}

\subsubsection{Computational Cost and Inference Speed}

Because interactive US simulation requires low-latency image generation, we evaluated the inference cost of each component of the pipeline on the same workstation. Runtime was measured with batch size 1 after GPU warm-up and is reported as mean time per image and frames per second (FPS). GPU memory was recorded as peak allocated memory during inference. We evaluated Pix2Pix, SDM with 750 and 1000 denoising steps, the SG-CycleGAN refinement stage, and the complete Pix2Pix+SG-CycleGAN two-stage pipeline.

Pix2Pix provided the fastest Stage~I inference because it is a feed-forward generator, reaching 155.25 FPS. SG-CycleGAN refinement was also fast when evaluated independently, reaching 130.54 FPS. The complete Pix2Pix+SG-CycleGAN pipeline achieved 63.33 FPS, indicating that this configuration is compatible with interactive or near-real-time simulation. In contrast, SDM required iterative denoising, resulting in substantially higher inference times of 27.87~s and 37.13~s per image for 750 and 1000 steps, respectively. As shown in Supplementary Fig.~2, increasing the number of SDM steps improved LPIPS up to 750--1000 steps, but also increased inference time substantially. We therefore selected 750 steps as a practical compromise between perceptual quality and computational cost. These results indicate that Pix2Pix is more suitable for low-latency or interactive applications, whereas SDM is preferable when anatomical fidelity is prioritized over generation speed.

\begin{table}[ht]
\caption{Inference time and GPU memory usage for Stage~I models, and Stage~II refinement. Runtime was measured with batch size 1 after GPU warm-up. FPS: frames per second.}
\label{tab:runtime}
\centering
\footnotesize
\begin{tabular}{lcccc}
\toprule
Model / Pipeline & Steps & Time per image (s) & FPS & Peak GPU memory (GB) \\
\midrule
SDM (Stage~I) & 750 & 27.87 & 0.036 & 0.15 \\
SDM (Stage~I) & 1000 & 37.13 & 0.027 & 0.15 \\
Pix2Pix (Stage~I) & -- & 0.0064 & 155.25 & 0.20 \\
SG-CycleGAN (Stage~II) & -- & 0.0077 & 130.54 & 0.13 \\
\bottomrule
\end{tabular}
\end{table}

\subsection{Qualitative Evaluation}

\subsubsection{Simple Scenarios}

\begin{figure*}[t]
     \centering
     \includegraphics[width=1\textwidth]{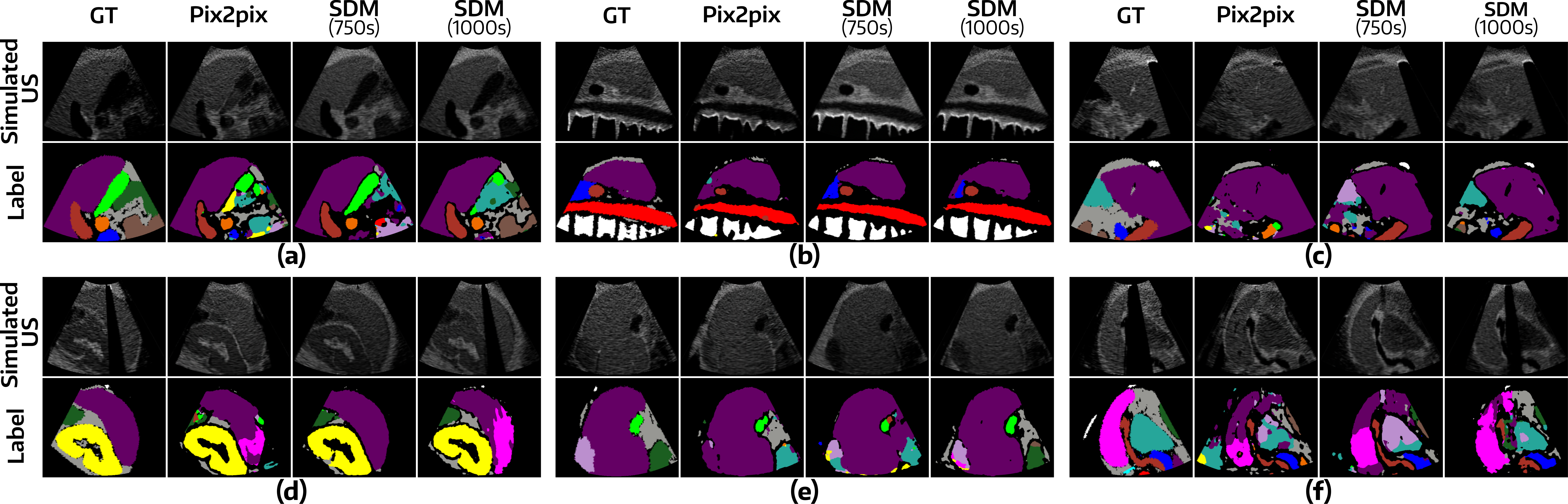}
     \caption{Comparison of simulated US images from the physical model and both SIS models. Top: US simulated scan. Bottom: predicted anatomical segmentations from Segmentator $S$. Anatomical colors follow the label convention described in Section~\ref{methods:segmentation}}
     \vspace*{2ex}
     \label{fig:results}
 \end{figure*}

We qualitatively assessed the scattering patterns, echogenicity, and anatomical fidelity of the US simulations produced by each Stage~I model. Figure~\ref{fig:results} show representative examples from six typical abdominal views, together with the corresponding anatomical segmentations predicted by the segmenter $S$. Overall, segmentation performance was consistent for larger structures such as the liver and kidneys, regardless of the simulation approach. However, SDM better preserved smaller tissues and structures with echogenicity similar to adjacent organs. In particular, SDM more accurately maintained the shape and echogenicity of organs near the liver—such as the pancreas, stomach, and tumor regions—whereas Pix2Pix tended to simulate these areas as part of the liver (see Fig.~\ref{fig:results}b,c,e). This difference is clearly reflected in the segmentations produced by $S$: in SDM-generated images, the pancreas (black), stomach (cyan), and tumor (light purple) are preserved as distinct labels, while in Pix2Pix-generated images these regions are frequently misclassified as liver (purple). 

Overall, SDM better preserves abdominal anatomical structures. However, smooth tissue interfaces and subtle echogenicity transitions still pose challenges, particularly for Pix2Pix. For example, in Fig.~\ref{fig:results}a, SDM correctly maintains the anechoic nature of the gallbladder, whereas Pix2Pix introduces unrealistic echoes within the area. This is also reflected in the segmentation produced by $S$, which misinterprets the region as a mixture of neighboring tissues rather than an homogeneous, anechoic area, as is typical of the gallbladder (green).

We also analyzed scenes with complex acoustic interactions, such as rib shadowing. Three of the example windows include ribs (Fig.~\ref{fig:results}c,d,f), which have high acoustic impedance and should produce bright reflections followed by posterior acoustic shadows. SDM (evaluated with both 750 and 1000 diffusion steps) was able to reconstruct the presence of a rib and its shadow in Fig.~\ref{fig:results}c. When the rib was located near the upper image boundary (Fig.~\ref{fig:results}d,f), the shadow was reproduced only by the model when using 1000 diffusion steps. This difference is clear in Fig.~\ref{fig:results}d: at 750 steps, SDM omits the shadow, and therefore $S$ labels the region as part of the liver. At 1000 steps, the shadow is actually produced but visually splits the organ, making it hard for the segmenter to recognize it as a single structure. Furthermore, although the image may look correct to the eye, the segmentation model struggles to separate the spleen (pink) from the liver (purple). Pix2Pix, on the other hand, fails to reproduce both the rib and its shadow in all cases.

In terms of speckle noise, the two methods produced noticeably different textures. Pix2Pix outputs more closely replicated the scattering pattern of the physical model, including the characteristic horizontally elongated speckle texture. SDM’s outputs, in contrast, had a more uniform and isotropic speckle pattern with less directional elongation.

\subsubsection{Pathological Scenarios}
\label{sec:Patho}

\begin{figure*}[t]
    \centering
    \includegraphics[width=1\textwidth]{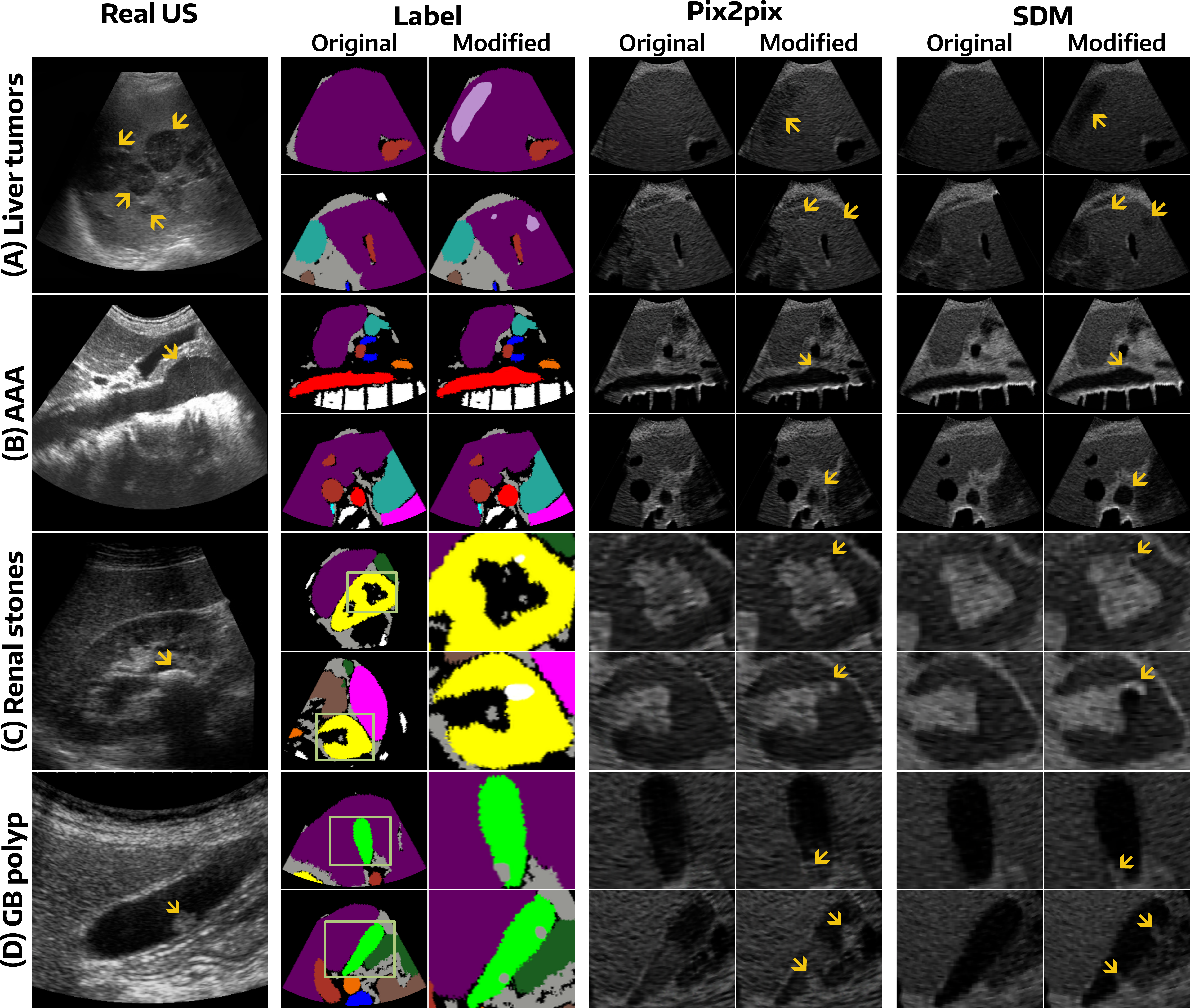}
    \caption{Simulation of pathological cases with SIS models. Each row shows a real US reference (left), original and modified label maps (center), and Pix2Pix- or SDM-generated images (right), with and without synthetic pathologies. Yellow arrows highlight the inserted pathological findings}
    \vspace*{2ex}
    \label{fig:deformation}
\end{figure*}

To further explore the controllability of the proposed label-based simulation framework, we generated proof-of-concept pathological scenarios by introducing anatomical alterations and abnormal structures into otherwise healthy anatomical maps. These experiments were intended to assess whether the models respond coherently to controlled label-space edits. In particular, we simulated synthetic liver tumors, abdominal aortic aneurysms (AAAs), kidney stones, and gallbladder polyps. Liver tumors were simulated by adding lesions of varying sizes and positions within the liver, using the existing tumor label; AAAs were created by thickening the aortic wall to mimic aneurysmal dilation in sagittal and axial views. Kidney stones and gallbladder polyps, on the other hand, were labeled as “bone” and “fat,” respectively, to match their acoustic behavior and US wave interactions~\cite{bushberg2011essential}. \textcolor {black}{Figure~\ref{fig:deformation} displays the results for these four cases together with real US images of the corresponding findings, included only as qualitative visual references}. Each row shows (from left to right) a real US reference image, the original and modified label map, and the corresponding outputs from Pix2Pix and SDM (with and without the synthetic pathology). 

Figure~\ref{fig:deformation}(A) presents hepatic tumors, which often appear as hypo- or hyperechoic lesions~\cite{hoener2020sonographic}. We simulated two tumors of different sizes and positions. Both SIS models generated a visible lesion for the larger tumor. For the smaller lesion, SDM better preserved the inserted structure, whereas Pix2Pix tended to blend it with surrounding tissue.

Figure~\ref{fig:deformation}(B) illustrates AAA, typically diagnosed when the aortic diameter exceeds 50\% of the proximal segment~\cite{pande2008abdominal}. In the sagittal view, both models reproduced the enlarged aortic region. In the axial view (bottom), SDM preserved the anechoic lumen, while Pix2Pix produced a distorted or absent aortic structure.

Kidney stones are highly echogenic and usually cast posterior shadows~\cite{simon2017some}. As shown in Fig.~\ref{fig:deformation}(C), both models generated hyperechoic responses for large synthetic stones, with SDM producing a thinner, localized artifact. Pix2Pix also captured the reflection but exaggerated the effect across the entire structure and failed to reproduce the expected acoustic shadow. For smaller stones, SDM preserved the reflective properties but generated limited shadowing, while Pix2Pix blended the stone with the renal pelvis.

Fig.~\ref{fig:deformation}(D) shows synthetic gallbladder polyps, which are intraluminal projections of the gallbladder wall, typically non-shadowing and hyperechoic~\cite{persley2005gallbladder}. We simulated polyps using the fat tissue label. Both models produced hyperechoic regions consistent with the label modification. However, Pix2Pix produced a diffuse hyperechoic effect, which blends with a poor representation of the gallbladder (top). In contrast, SDM provided clearer polyp-like artifacts with better spatial definition and anatomical alignment.

\subsubsection{Stage II: Image Refinement (Qualitative)}

\begin{figure*}[t]
    \centering
    \includegraphics[width=1\textwidth]{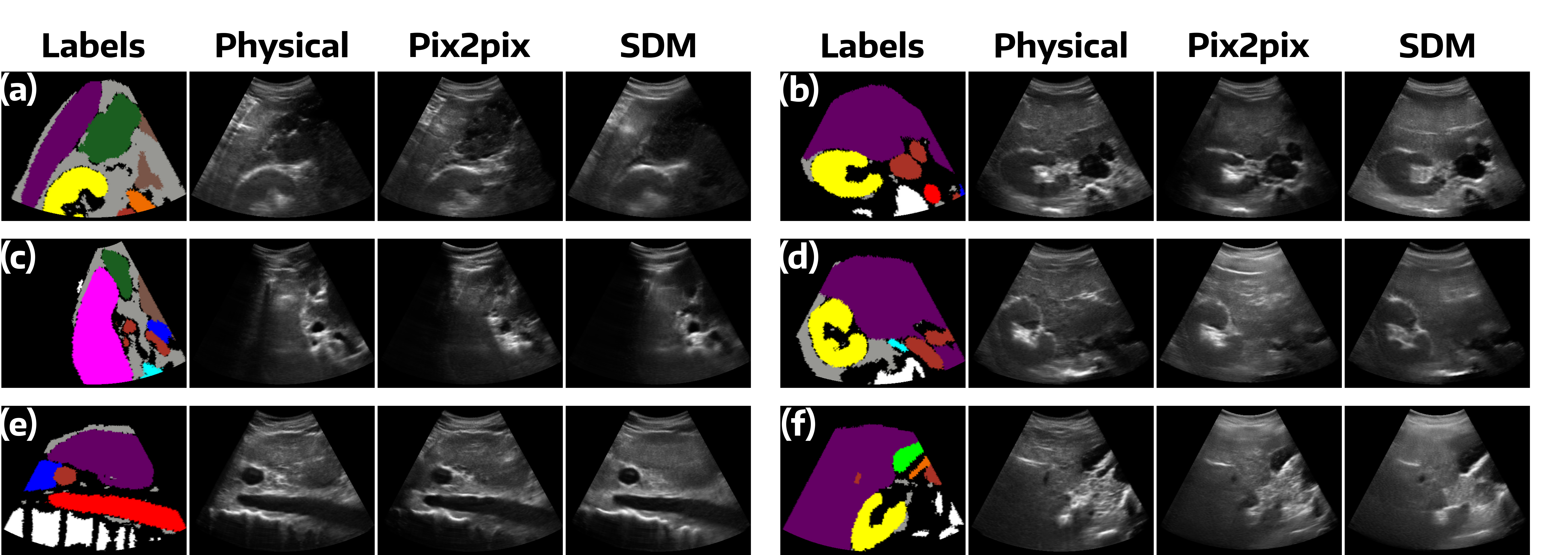}
    \caption{Refinement results from SG-CycleGAN using different inputs: physical model (GT), Pix2Pix, and SDM. First column shows the reference label maps. Anatomical colors follow the label convention described in Section~\ref{methods:segmentation} }
    \vspace*{2ex}
    \label{fig:refinement}
\end{figure*}

Figure~\ref{fig:refinement} illustrates the outputs refined by SG-CycleGAN when using inputs from the physical model, Pix2Pix, and SDM for the same representative slices. The purpose of this qualitative analysis was to examine not only whether refinement improved visual realism, but also whether it introduced visible hallucinations, removed anatomical information, or altered relevant image structures.
Across the representative cases, SG-CycleGAN mainly modified texture, contrast, and speckle appearance, while preserving the coarse anatomical layout inherited from the input image. Major structures such as the liver, kidney, gallbladder, vascular regions, and visible abdominal interfaces remained spatially consistent after refinement. We did not observe gross hallucinated organs or complete removal of major anatomical structures in these examples. 
However, the refinement stage was not anatomically neutral. Local texture and contrast changed after translation, and these changes depended on the Stage~I input. Images refined from Pix2Pix inputs most closely matched the speckle texture and detailed appearance of the physical-model refined outputs, likely because Pix2Pix already generated texture statistics closer to the physical-model domain. In contrast, SDM-derived images remained smoother after refinement and showed more homogeneous intra-organ scattering, especially in large structures such as the liver. This smoothing reduced some exaggerated hyperechoic lines produced by the physical model, but it could also attenuate fine textural details and small local boundaries.
 
\section{Discussion}
\label{sec:Discussion}

\textcolor{black}{This work addresses a central limitation of many abdominal US simulation pipelines by investigating whether the CT-based ray-casting stage can be approximated by generative models operating directly from semantic anatomical labels. The proposed approach does not eliminate the need for CT-derived information during training, since ray-casting simulations remain necessary to provide paired supervision for Stage~I. However, once trained, the generative model can synthesize a simplified US image directly from semantic labels without requiring a new patient-specific CT or MRI volume or re-executing the ray-casting procedure. This preserves anatomical and acoustic priors learned from physics-based supervision while providing greater flexibility at inference time. In practice, the required semantic maps may originate from reusable anatomical templates, atlas-based models, manually edited cases, procedural anatomy generation, or deformations of existing labeled anatomies.}

\textcolor{black}{The Stage~I comparison revealed a clear trade-off between morphological consistency and perceptual similarity. SDM better preserved spatial and anatomical structure, whereas Pix2Pix more closely matched the perceptual and distributional characteristics of the physics-based reference domain. These metric groups capture complementary properties. Morphological metrics reflect reconstruction fidelity and preservation of anatomical structure, whereas perceptual metrics characterize feature-level and distributional similarity. This distinction is particularly important in medical image synthesis, where a visually plausible image is not necessarily anatomically correct~\citep{kazerouni2023diffusion,skandarani2023gans,kocak2025evaluation}. Moreover, perceptual metrics such as LPIPS, FID, and KID do not directly assess anatomical correctness or ultrasound-specific acoustic consistency and should therefore be interpreted together with anatomy-aware and qualitative evaluations. The relatively low absolute SSIM values observed for both models should also be interpreted in this context. Semantic maps encode organ geometry but not the fine-scale acoustic texture, speckle realization, or local scattering variations of the corresponding simulated US image. Consequently, small local differences in luminance, contrast, and texture can reduce SSIM even when the main anatomical structures remain consistent. SSIM should therefore be interpreted primarily as a relative measure of structural agreement in this setting.}
\textcolor{black}{The stronger morphological performance of SDM is consistent with semantically conditioned diffusion models, whose iterative denoising process repeatedly incorporates spatial information from the label map and can favor preservation of organ geometry, small structures, and controlled label-driven modifications~\citep{wang2022semantic,stojanovski2023echo,dominguez2024diffusion}. However, iterative denoising may also smooth high-frequency US texture and reduce local speckle variability, which could explain the weaker perceptual and distributional scores. Pix2Pix exhibits the complementary behavior. Adversarial training encourages high-frequency detail and sharp local texture~\cite{isola2017image,zhu2017unpaired,skandarani2023gans}, which is consistent with its closer perceptual match to the physics-based reference domain and with previous GAN-based US synthesis approaches~\cite{Pigeau,zhang2020deep,liang2022sketch}. At the same time, adversarial objectives do not explicitly guarantee preservation of small or uncommon anatomical structures, which may explain the weaker morphological consistency observed for Pix2Pix.}
\textcolor{black}{From a practical perspective, the choice between SDM and Pix2Pix depends on the intended application. SDM is preferable when anatomical and morphological consistency is prioritized, for example in controlled semantic scenario generation or offline synthetic dataset creation. Pix2Pix, in contrast, provides substantially lower inference latency and better perceptual and distributional similarity, making it more suitable for latency-sensitive applications such as interactive or real-time simulation. Thus, the two models are retained as alternative Stage~I implementations rather than as parallel components that must be executed together. Neither represents a universally optimal solution; instead, they provide different operating points in the trade-off between anatomical fidelity, perceptual appearance, and computational cost.}

\textcolor{black}{Stage~II uses the previously established SG-CycleGAN refinement strategy~\cite{vitaleimproving} as a downstream compatibility test for the new Stage~I outputs. Both Stage~I alternatives could be processed by the refiner and shifted toward the real-US appearance domain. The segmentation-guided loss helps constrain the unpaired translation and preserve the anatomical organization established in Stage~I, while the adversarial component mainly modifies texture, speckle appearance, scattering, and contrast. However, refinement effectiveness still depends on how closely the Stage~I output matches the source distribution used to train SG-CycleGAN. Pix2Pix outputs were more compatible with this distribution, whereas the smoother SDM outputs showed a larger domain shift. Thus, Stage~II can improve visual realism but cannot fully compensate for differences introduced during Stage~I generation.}

\textcolor{black}{The choice of Stage~I model had a smaller effect on organ-level echogenicity after refinement. Similar $\chi^2$ values for Pix2Pix- and SDM-refined images suggest that SG-CycleGAN can normalize regional intensity distributions even when the input texture differs. This also shows that similarity to the physics-based simulation domain is not necessarily equivalent to similarity to real US data. Nevertheless, Stage~II should not be considered anatomically neutral. Although no gross hallucination or removal of major structures was observed qualitatively, adversarial refinement may still alter local texture, contrast, boundaries, or small findings, and dedicated expert assessment would be required to exclude subtle clinically relevant changes.}

The ability to manipulate the label space provides a practical mechanism for controlled scenario generation. Unlike CT- or MRI-dependent simulation, where anatomical variability is constrained by the available volumes, semantic labels can be directly modified to introduce deformations, motion-like transformations, or synthetic pathological findings. In our experiments, liver tumors, abdominal aortic aneurysms, kidney stones, and gallbladder polyps produced qualitatively plausible US appearances when represented through existing or acoustically analogous labels. These results extend the use of dense semantic conditioning such as SPADE and SDM~\citep{park2019semantic,wang2022semantic} to multi-organ abdominal US simulation and suggest potential applications in rare-case generation, synthetic dataset construction, and simulator case authoring. However, these pathological examples should be interpreted as proof-of-concept scenarios rather than clinically validated disease models.\textcolor{black}{Such controllability may also complement educational US simulators by expanding anatomical and pathological case diversity~\citep{ostergaard2016simulation,nayahangan2021simulation,dietrich2022ultrasound,lucius2023use}. Nevertheless, these remain potential applications. The present study does not include an educational intervention, expert reader assessment, or transfer-to-real-scanning evaluation.}

Several limitations should be considered. First, although inference does not require CT or MRI, training still depends on anatomically plausible segmentations and ray-casting simulations. Errors from automatic segmentation, simplified acoustic parameter assignment, or the physics-based simulator may therefore propagate into the learned models. \textcolor{black}{The study also contains a relatively small number of CT volumes, limiting the anatomical and pathological variability represented during training and evaluation. Similar cohort sizes have been used in related US synthesis studies~\cite{Tomar2021,dominguez2024diffusion}, but larger and more diverse populations remain necessary to assess robustness and generalizability.}

Second, the semantic labels provide a simplified representation of tissue acoustics and cannot fully capture intra-organ heterogeneity, anisotropy, microstructure, vascularity, probe pressure, or patient-specific acquisition conditions. \textcolor{black}{Third, the pathological cases were evaluated only as qualitative proof-of-concept examples using real pathological US images as visual references. Their correspondence with real pathological appearances has not been quantitatively or clinically validated and will require dedicated pathological datasets and expert assessment.}

Finally, the realism and generalizability of the refined images depend on the Stage~II real-US domain. SG-CycleGAN~\cite{vitaleimproving} was retrained using data from 11 healthy volunteers acquired with a single scanner and protocol. \textcolor{black}{Consequently, the current refinement stage may not capture the full variability associated with body habitus, scanner manufacturers, institutions, acquisition protocols, operators, or pathological conditions. Because the real-US refinement domain contains only healthy volunteers, the Stage~II evaluation primarily reflects similarity to healthy US appearances and cannot establish pathological realism.}

Future work should focus on broader datasets, model acceleration, and external validation. Larger multi-center, multi-scanner, and pathology-rich US datasets are needed to improve generalizability. Alternative conditional generative approaches, including ControlNet-like diffusion models, VAEs, and flow-based methods, should also be investigated as larger abdominal datasets and ultrasound-specific pretrained models become available. Additionally, diffusion acceleration and distillation may reduce inference time and facilitate interactive applications~\citep{nichol2021improved,kazerouni2023diffusion}.

\textcolor{black}{The present study should be interpreted as a technical feasibility assessment rather than as evidence of clinical realism or educational effectiveness. Dedicated expert assessment is still required to determine whether Stage~II refinement introduces subtle hallucinations, removes small pathological findings, or alters clinically relevant acoustic artifacts. Future validation should therefore combine perceptual and distributional metrics with anatomy-aware measures, acoustic artifact assessment, blinded expert review, pathology-specific evaluation, and task-based studies with sonographers or trainees. Downstream experiments should also determine whether synthetic data improve segmentation, detection, or acquisition-skill learning.}


\section*{Declarations}
\small \noindent \textbf{Author Contributions:} Santiago Vitale (S.V), Duilio.E Deangeli (D.E.D), Jose.I Orlando (J.I.O), and Ignacio Larrabide (I.L) contributed to the conceptualization and design of the study. S.V. and D.E.D. developed, implemented, and validated the deep learning models. S.V. and D.E.D. performed the experiments and drafted the manuscript. J.I.O. coordinated the project. J.I.O. and I.L. jointly supervised the work. All authors contributed to the interpretation of the results, critically revised the manuscript, and approved the final version.\\ \\
{\small \noindent \textbf{Funding:} This work is funded by PIP GI 2021-2023-11220200102472CO. A NVIDIA GPU hardware grant supported this research with the donation of a NVIDIA Quadro P6000 GPU.\\ \\
{\small \noindent \textbf{Code and Data Availability:} 
\textcolor{black}{The datasets used in this study are publicly available through the VISCERAL Anatomy3 Challenge and IRCAD repositories, subject to their respective terms of use. The processed dataset generated for this work is publicly available at \url{https://www.kaggle.com/datasets/ignaciorlando/ussimandsegm}. Additional reproducibility resources are available at \url{https://github.com/thepochynsons/semantic-label-abdominal-us-synthesis}, including representative semantic inputs, test samples, Pix2Pix and SDM outputs, data-organization and preprocessing information, and instructions for reproducing the experimental and evaluation procedures. The complete source code cannot be publicly released due to project-related restrictions. Additional implementation details and materials are available from the corresponding author upon reasonable request.}}\\ \\
\noindent \textbf{Compliance with ethical standards.} This article does not contain any studies with human participants or animals performed by any of the authors. \\ \\
\noindent \textbf{Conflict of Interest:} The authors have no relevant financial or non-financial interests to disclose.\\ \\
\noindent \textbf{Informed consent: } Not applicable\par}

\bibliography{sn-bibliography}

\end{document}